# How can technology improve acetabular and pelvic fractures management in the operating room? A systematic review

Running title: New technologies in pelvic fractures

## Author informations:

Marie Le Baron[1,3*], Martine Pithioux[2, 3], Stéphane Candoni[3], Xavier Flecher[1,2,3]

[1]Department of Orthopedics and Traumatology, Institute of Movement and Locomotion, CHU Nord, Marseille, France

[2]Department of Orthopedics and Traumatology, Institute of Movement and Locomotion, Sainte-Marguerite hospital, Marseille, France

[3]Aix Marseille University, CNRS, ISM, Marseille, France

*Corresponding author:
marie.le-baron@univ-amu.fr
ORCID: 0000-0002-8833-7454
Address: Service de chirurgie orthopédique, Hôpital Nord, chemin des Bourrely, 13015 Marseille, France
Tel: (+33) 0491968694
Fax: (+33) 0491964977

Disclosure: all authors listed meet the authorship criteria according to the latest guidelines of the International Committee of Medical Journal Editors, and all authors are in agreement with the manuscript.

Acknowledgments: the authors have nothing to report.

Conflicts of interest: the authors declare no conflicts of interest.

Author contributions:
Conceptualization: MLB, XF; literature search and data analysis: MLB, MP, SC, XF; writing original draft: MLB; writing review and editing: MLB, MP, SC, XF.

## Abstract

Objectives: In recent years, there has been a proliferation of literature describing new technologies increasingly integrated into the surgical management of pelvic fractures. However, the available data are heterogeneous, scattered, and sometimes limited to small series or single-center studies. This review aims to present new technologies in the field of pelvic and acetabular fractures and explore their potential benefits for surgeons and patients. The goal is to describe current concepts, advantages, and limitations and to open a discussion about future use and development.

Methods: The search was conducted through Medline, Central and Embase databases from inception to January 2025 following the PRISMA (Preferred Reporting Items for Systematic Reviews and Meta-Analyses) protocol.

Results: A total of 841 records were reviewed and 115 studies were included according to the inclusion and exclusion criteria. Technologies such as robotics, navigation, patient-specific management, and 3D printing are being developed for the management of pelvic injuries. These technologies demonstrated greater surgical accuracy without increasing complications.

Conclusion: In pelvic and acetabular fractures surgery, new technologies are very attractive and are a promising help for management of these complex injuries and need to be improved before being included in routine care.

## Statements and Declarations:

Competing interests:
XF: educational consultant for Zimmer Biomet, Stryker endoscopy
MLB: deputy editor for European Journal of Orthopaedic Surgery & Traumatology
MP, SC: no competing interests

Funding declaration:
No funding was received to assist with the preparation of this manuscript.

## Abbreviations

CAS: Computer-assisted surgery
RAFR: robot-assisted fracture reduction
APC: anteroposterior compression
LC: lateral compression
VS: vertical shear
RA: robot assistance
SI: sacro-iliac
PS: patient specific

## Introduction

Although they are rare, pelvic and acetabular fractures remain challenging for orthopaedic surgeons. These fractures may occur either in a young and active population with high-energy trauma and other injuries, leading to a reduced level of previous activities, or in an elderly and frail population, where maintaining autonomy is crucial. In both cases, fracture pattern understanding, choice of approach and method of fixation remain controversial. Nevertheless, due to the complex anatomy of the pelvic bone, surrounded by nerves, vessels and visceral structures, reduction and fixation remain difficult even for expert surgeons, leading to an increased risk of complications, as malposition of hardware (1) or soft tissues lesions.

Traditionally, open reduction and internal fixation have been considered as the gold standard, but less morbidity and good outcomes with mini-invasive surgery led to a development of these techniques in this setting.

In order to prevent these potential complications, the use of new technologies such as computer-assisted surgery (CAS) (2) is becoming increasingly common, given the good results achieved in other fields: robot for joint implant positioning, navigation in spinal surgery. CAS includes preoperative planning, intraoperative imaging and navigation (3). An alternative can be personalization: 3D printing and patient case planning focus on the patient and their fracture, rather than just their fracture pattern.

This review aims to present some new technologies in the particular field of pelvic and acetabular fractures, exploring their potential benefits for surgeons and patients. It was not the purpose of this systematic review to develop guidelines for specific indications, but rather to report the results in a summarized manner based on what has been published in the literature. The objective is to describe their current concepts, their advantages and limitations, and initiate a discussion about their future use and development.

## Material and method

This systematic review was conducted in accordance with the PRISMA (Preferred Reporting Items for Systematic Reviews and Meta-Analyses) guidelines (4). The aim was to identify and analyze studies evaluating innovative technologies to help fracture reduction or fixation of pelvic or acetabular fractures.

A comprehensive literature search was carried out in the Medline, Central and Embase databases.

Keywords (Medical Subject Heading (MeSH) terms) used combined using Boolean operators included: “pelv* fracture” OR “acetabul* fracture” AND “patient specific” OR “navigation” OR “robot” OR “3D printing”. This review was not registered on a systematic review database. The search strategies were performed in January 2025.

Exclusion criteria included studies without pelvic fracture or surgery, children <18 yo, studies evaluating outcomes of visceral lesions or urologic lesions and not bone injuries, periprosthetic or pathological fractures, vertebral fractures, lumbo-iliac fixation, malunion or nonunion, studies conducted on animals, reviews, editorials and conference abstracts, other languages than English and full text not available. No restrictions were placed on date of publication. All studies were considered for eligibility. Both titles and abstracts of studies were screened. Duplicate articles were removed. The full texts were obtained and were reviewed. Details are described in the flow-chart (Figure 1). To be eligible for inclusion, a publication had to present information on the use of new technologies in managing pelvic fractures. This information had to include details on surgical outcomes, surgical indications, or descriptions of procedures and any associated complications (such as those related to patient specific, navigation, robot or 3D printing). Publications that did not meet these inclusion criteria were excluded. Additionally, the study had to focus primarily on the use of new technologies in managing pelvic fractures. If the publication only mentioned the technologies in passing without focusing on them, it was excluded.

## 1. Technologic tools

### - Robotic systems

**Introduction**

A robot consists of a computer with a surgical planning software, an optical tracking system and a mechanical arm with navigation tools, in association or not with a 3D fluoroscopy system. At the beginning of the procedure, the robot takes and analyzes images of the patient with landmarks on the patient to prepare the guidance. The surgeon can plan the surgical procedure before or during the procedure. The robot with the mechanic arm will guide the surgeon according to the plan during the operation. It is expected that the robot will act in an ideal, anatomical way, as planned. In the pelvic surgical field, the plan is decided to reconstruct the fracture area by mirroring to the healthy side (5). Thus, its application is limited to unilateral fractures.

Most of these studies apply the use of the robot for pelvis fractures. Few robots are used and almost only Chinese studies are available. The most used is TianJi Robot (Ti-robot=TinaviTM orthopedic surgical robot system (TINAVI Medical Technologies, Beijing, China (6–8)). Another robot described in Zhao's study, is the UR16e robotic arm (9).

**Help for the reduction**

A first experimental study of robot-assisted fracture reduction (RAFR) developed a 3D-based image-guided serial robot for pelvic fracture reduction, with manual and semi-automatic reduction pathways (10). Some studies have developed algorithms that automatically find a safe pathway to achieve an effective pelvic fracture reduction (11). Ideally the minimal movement is required to avoid adding soft tissues lesions. These algorithms take account of pelvic fracture displacement, muscle forces, and avoiding collisions between bone fragments and organs during reduction. They have been developed on pelvic and not acetabular fractures models (APC2, APC3, LC2, LC3, VS) (11,12). Reduction can be done manually using Shanz pins placed in the injured side to allow its movement (13), but need strong maneuvers consuming strength from the surgeon (14) and outcomes can't be guaranteed (10). In addition to the already described closed reduction frames (Starr frame (15), UCRT frame(16)), biomechanical studies try to develop the accuracy of the “reduction tools” (13,17). Clinical applications have already started and seem to show satisfactory outcomes in both young and older patients (9,18,19).

**Help for the fixation**

The aim of using the robot is to perform positioning and placement of hardware with the entry point, trajectory and exit point of a screw for example, as planned. This has been described both in acetabular fractures (20) and pelvis models fractures (21,22). Another application is to perform the tissues dissection with the robot (Da Vinci, used for visceral, thoracic or urologic surgery) before anterior plate fixation of an anterior column and posterior hemitransverse acetabular fracture to decrease soft tissues trauma associated to an open approach (23).

**Results**

First, regarding reduction: literature showed few results about RAFR on pelvic fracture, with good outcomes reported. Comparison between these results is difficult according to different populations and different outcomes. Only three studies were found to describe outcomes of RAFR on pelvic fractures (9,19,24). Results are shown in the table 1.

Second, regarding fixation: were found one retrospective study (25) and 9 comparative studies on fracture fixation with RA (6,8,26–32), one of which was prospective without randomization (28). These comparative studies included from 21 to 119 patients, with only SI screws or posterior and anterior pelvic ring screws. They showed shorter operative time and lower intra-operative radiation exposure. RA allowed a good accuracy of screws placement. The quality of reduction (according to the Matta score) was better or not inferior with RA than with the conventional method. Functional results (Majeed score) were similar with both methods. Results are described in Table 2.

**Limitations**

Whether for reduction or fixation of pelvic fractures, the robot requires the reference on the healthy pelvis (24) to act on the injured side, and the path for achieving the reduction is based on mirroring: it limits his use to non-bilateral injuries. Moreover, in case of comminution of iliums, it could be difficult or impossible to place pins.

This equipment represents an additional cost (27) for the hospital and superior to the conventional methods. Zhu (6) showed that the medical expenses were significantly higher in the RA group (32461.1 ± 3857.3 yuan) than in the conventional group (23543.2 ± 5341.6 yuan) ($p<0.05$), but not than the ORIF group (32340.8±4892.0 yuan).

No clinical studies on acetabular fractures showing outcomes were found.

## - Navigation without robotic assistance: computer-aided navigation systems

**Introduction**

Navigation used in medical fields includes ultrasound, electro-magnetic or electro-optic guidance. Navigation in surgery is mainly electro-optic and has been developed specifically for neurosurgery (33). It has also been developed in orthopedic surgery and trauma, to ensure precision of hardware positioning, to decrease operative radiation time and to allow mini-invasive surgery (34,35). A tracking system is connected to an imagery modality, fluoroscopy, CT or MRI. Registration of the operative zone, of the instruments, and of imaging of the operative area is required. Few systems exist: OARM Medtronic / AIRO mobile / Stryker Navigation System II-Cart, Kalamazoo, MI, USA + Ziehm Vision FD Vario 3D (Ziehm Imaging, Nuremberg, Germany).

The conventional tracking system is an invasive device. Rosenberger (36) described a new type of registration device, noninvasive, in some cases of acetabular and pelvic surgeries. It uses a double vacuum system plus cushions as a noninvasive immobilization system. Targeting device is hold on the immobilization system, avoiding an invasive placement.

Lehmann (37) showed in a cadaver study and Pishnamaz (38) in an experimental study the use of electro-magnetic navigation system feasible placement of screws for acetabular fractures. This technic doesn't need registration of instruments or operative zone and theoretically saves time during surgical procedure while providing less intraoperative radiation exposure. This allows real time navigation, unlike other methods. One limitation remains some potential interferences on the electromagnetic field.

**Help for the reduction**

Navigation is rarely used for fracture reduction and is only used for pelvic fractures. Li (39) combined a mixed reality surgical navigation with a frame to perform a closed reduction, allowing a real-time monitoring of fracture fragment displacements. In another study, from preoperative CT and intraoperative fluoroscopy, Han (40) used a multi-body registration framework for reduction planning in some cases of fracture-dislocation.

**Help for the fixation**

Navigation guidance of percutaneous screws insertion can be performed with fluoroscopy (41), either on pelvis and acetabular fractures, with an accurate placement. Takao (42) described a technique with fluoroscopy guiding matching preoperative CT scan. The main concern was the discordance between the preoperative images and the peroperative reality after fracture reduction. Attention was on the contralateral sacral fragment during image registration, so the navigation system could track his movement even after fracture reduction. Han (40) developed a technic to match the preoperative CT scan of the patient to a pelvic model constructed from 40 pelvic CT scans. Navigation can also be performed with intraoperative CT guidance (43,44), derived from the neurosurgery, like in pedicle screws placement. In pelvic fractures fixation, a prototype system for freehand drill guidance to support the surgeon to position the K-wire at the bone surface was presented in a first preclinical study (45).

Some studies have described percutaneous fixation under CT guided navigation for non-displaced acetabular fractures: in 4 cases of coronal roof acetabular fractures (46) (2 fragments, non-intraarticular step), and in fractures involving quadrilateral plate (47) (5 patients, in a percutaneous way). Navigation may also help safe placement of screws by an open approach close to the joint (48), like in fractures with an impaction.

**Results**

Nine clinical studies (49–57) described the results of pelvic fractures fixation under navigation. Five were non-comparative (49,51,54,55,57), and 4 comparative (50,52,53,56), one of which was prospective (53). Most of non-comparative studies demonstrated the accuracy of screw placement under navigation. Comparative studies found reduced operative time, fluoroscopy duration and blood loss with navigation. The prospective multicenter study (53) found no statistical differences between two groups regarding intraoperative or postoperative complications and more neurological complications in the navigated group. Results are described in Table 3. A comparative experimental study (58) with 210 screws inserted in 15 sawbones (30 hemipelvises) and 15 human cadaver specimens (30 hemipelvises) between screw placement under navigation and conventional technique found a lower perforation rate under navigation (7% vs 20%, significant), with a more accurate placement. Analyzing by type of screw, the posterior column screws placement was improved by navigation, unlike the anterior column screw placement. However, this study showed that fluoroscopy-based navigation procedures led to significantly increased operative and fluoroscopic times and radiations doses.

Two studies (59,60) focused on procedures with pelvic ring fixation in a hybrid OR, defined by an operating room including a fixed imaging system, either a combination of a surgical environment with advanced intraoperative imaging. These two studies showed a good accuracy of screw placement in this environment. Richter (59) studied SI and TSTI (transsacral transiliac) screws placement (210 screws in 187 patients). He found that 90.6 % of SI screws and 80.7 % of TSTI screws showed no cortical perforation.

In the other study (60), 41 patients with fragility fractures of the pelvis underwent 121 percutaneous screws. Mean operative time was 84 ± 38.7 minutes, and the mean blood loss was 7.6 ± 3.8 g. There were 119/121 screws with no misplacement.

Three studies (61–63) described the results of navigation in acetabular fractures management, of whom one was comparative and two were non-comparatives. The comparative study (61) found a decreased operative and radiation times, and a decreased estimated blood loss in the navigation group. Additionally, navigation seems to provide a good accuracy of screw placement. Results are described in Table 4.

**Limitations**

Again, Additional cost for this equipment remains, as the robot system, a concern. It is estimated at more than 300,000 USD (63).

## 2. Personalization

- <u>Patient specific</u>

The use of patient specific (PS) hardware is based on pelvic symmetry: the healthy side is used as a reference model to adapt the device to the fractured side by mirroring (64). This study included 14 healthy pelvises, right sides being reflected and aligned with the left sides. No significant difference was found in terms of volume and surface between two sides. Deviation analysis showed that more than 90% of measures were <2mm.

**Help for reduction planning**

Reduction planning was proposed using preoperative CT images (bidirectional framework in pelvic fractures (65), multibody registration framework (40) in both pelvic and acetabular fractures, tested with 3 cases in a cadaveric study).

In acetabular fractures cases, Fornaro (66) described the use of an interactive planning tool with a haptic device that generas haptic feedback, to develop a preoperative PS virtual reduction planning in 7 patients. A good correlation has been demonstrated between the plan and the postoperative CT-scan. Boudissa (67–69) conducted 3 studies on virtual reduction with a PS biomechanical model simulator. This system is able to simulate the action of forces on the fragments and interactions between soft tissues and bone. These studies showed a good quality of the reduction: anatomic for 12/ 14 patients (67), anatomical reduction in 90% in simulation group (9/10) vs 60% (12/20) (non-significant) (68) (case control study) and better reduction in the simulation group (poor reduction 7.7% (2/26) versus 27.5% (11/40) $p = 0.048$) (69) (prospective non-randomized comparative study). Results showed also a reduced mean operative time in the simulation group (113 min ± 33 (60–180) versus 196 min ± 32 (60–260) ($p = 0.01$) (68), 110 minutes vs. 155 minutes, $p = 0.01$, (69)) and a reduced mean blood loss (505 mL ± 189 (100–750) versus 745 mL ± 130 (200–850) ($p < 0.01$) (68), 420 vs. 670 mL, $p = 0.01$) (69).

**Help for the fixation**

In relation to 3D printing technology (cf next part), some studies have shown how to design patient-specific templates to guide hardware placement in pelvic fractures. For instance, there are templates allowing insertion of pins on iliac crest for external fixation (70) or iliosacral fixation in pelvic ring fractures, positioned on the pins of the external fixator placed on previous surgery (71).

This technic is also used in acetabular surgery, by developing Patient Specific Instruments (PSI) such as PS drilling guides associated with PS implants (contoured plates) (72). Another application in acetabular fractures involves preparing some plates before surgery by precontouring them based on the anatomy of the other side, after mirroring (73). Augmented reality could be incorporated in this process to assist with designing and contouring PS plates (74).

**Results**

Clinical results are available for acetabular fractures. Five clinical studies (73,75–78) were found, 2 non-comparative (73,78) and 3 comparative (75–77), one of which was a prospective randomized study with 21 patients (75). This study showed a decreased operative time and blood loss in the PS group, with a good reduction. Results are described in Table 5.

**Limitations**

Reduction planning includes a specific and additional preoperative time for each case, including a preoperative semi-automatic segmentation time (from 82 min ± 18 (60–120) to 156 min ± 37.9 [120–180]) and a preoperative simulation time (from 22 min ± 4 (18–31) to 23 min ± 9 [16–38]) as well as the time required to learn how to use the system (67–69). Similarly, precontouring the plate may reduce the operative time, but it requires additional preoperative time.

## - <u>3D printing</u>

**Introduction**

The 3D printing process is step-by-step. First, images acquisition is required. This is mainly obtained from CT performed during trauma assessment. Next, the creation of a virtual model is performed; finally, the fabrication of a solid model is realized, as an exact reproduction of the patient's anatomy. This technique can be used for preoperative planning, surgical guide and implant fabrication, or surgical simulation.

**Help for fixation**

This can be used for several applications. Since the anatomy of the pelvis is complex and the hardware must be contoured to fit the bone and provide strong fixation for all types of fractures, some authors use 3D-printed pelvis based on a CT scan of the patient to pre-contour the plates before surgery for acetabular fractures. This method also aims to save time and reduce blood loss during surgery. This technic involves mirroring the healthy hemipelvis to work on the "reduced" injured side, with both anterior and posterior plates (Chana Rodriguez in a case report (79), then on a serie of 20 patients (80), Upex (81), Zou (82) in combination with para-rectus approach only). This technic can be combined with computer assisted surgery to perform a virtual reduction of the acetabular fracture, a virtual positioning of hardware, then 3D printing, and pre-contouring on solid 3D model, checking with postoperative CTscan (Zeng in a preliminary report of 38 patients (83), with minimal invasive para-rectus approach, and then in a clinical study of 10 patients (84)).

Designing hardware with the help of 3D printing technology is also described. Some anatomic posterior column plates were designed and printed preoperative CT scans (85): no recontouring was needed during surgery. In an experimental study (86), 3D printed plates led to a better reduction for acetabular fractures. Wang (87) reported the use of a 3D metal printed plate in a patient with an accurate positioning and reduction of the fracture, whereas Robinson (88) observed malposition of the hardware in more complex fractures (more than 5 fragments). A comparative experimental study (89) between virtual reduction and fabrication of a 3D printed plate and manual bended plate showed a best contact between plate and bone and a better reduction of fracture in the 3D printed group. In addition, a biomechanical study showed that 3D printed acetabular wing-plates provide same stability as conventional fixation systems (90).

3D printing can also help for positioning the hardware by planning surgery or designing guides to accurate placement, mainly for pelvic fractures. Preoperative planning of screw insertion for unstable pelvic fractures has been shown to decrease operation time and fluoroscopy time, without affecting the quality of reduction (91). 3D printed templates were designed to guide the entry points and screws trajectory in pelvic fractures: for antegrade trans-pubic screws, ilio-sacral fixation, pin insertion for external fixators on iliac crests and acetabular fractures fixation (70,71,92–95). In a cadaveric study, Tao (96) designed a 3D guide that is applied to the skin, not the bone (markers are attached to the skin before the CT scan), for anterior and posterior fixation of pelvic fractures.

**Results**

There are many applications for 3D printing, and many studies have shown positive results. Nine clinical studies (77,84,97–103) were found (7 comparatives (77,97,99–103) and 2 non-comparatives (84,98)), describing reduction and pre-contouring plates for acetabular fractures outcomes. Two of them were randomized (97,103). The studies concluded that reducing acetabular fractures and using pre-contoured plates on 3D-printed models decreases operative time, blood loss, and fluoroscopic time. The results are described in Table 6. Another comparative study (104) considered both pelvic and acetabular fractures, showing similar but not significant results.

The association of 3D printing with other technologies has been scarcely studied. A study of 3D printing combined with robot-assisted screw placement versus robot alone in pelvic fractures found shorter intraoperative screw placement time and greater screw positioning accuracy (22). Conversely, 3D printing and manual reduction versus 3D mirroring and virtual reduction for acetabular fractures demonstrated improved outcomes with the mirroring approach (105,106). In both studies, the 3D mirroring group had a shorter operation time, less blood loss, fewer fluoroscopic screenings, and fewer complications.

The 3D printing included 3D-printed hardware. Two studies examined the outcomes of acetabular fractures (107,108). The instrumentation time, surgical time, and intraoperative blood loss were shorter. No difference was found in the quality of reduction or the complication rate.

Another application of 3D printing is designing templates to position hardware. Three studies (92–94) examined the outcomes of using a 3D-printed template for screw placement in pelvic fractures. A comparative study (94) found that operative time was lower in the template group. Two non-comparative studies (78,95) described using a 3D-printed template for acetabular fracture surgery. Both studies achieved reduction of the fracture and accurate hardware placement. The results are described in Table 7.

**Limitations**

The time required for 3D printing remains a concern during the preoperative period, as the management of these fractures becomes increasingly early (98), primarily in sacroiliac disjunctions. However, two recent studies (109,110) showed that the use of 3D-printed surgical assistance had no effect on length of stay for two different populations: non-geriatric and geriatric patients. Only one study (80) addressed the time it takes to print a model for acetabular fractures: 385 minutes (range 322–539) and 238 grams of plastic (range 180–410).

## Conclusion

These new technologies may improve the management of pelvic and acetabular fractures. These promising systems include technological tools and personalization. Robotic systems demonstrated accurate hardware placement without increasing the complication rate. Intraoperative time and radiation exposure were reduced. These systems reported good reduction quality, though no comparative study has been performed. Most studies have shown that computer-aided navigation systems increase the accuracy of hardware placement and decrease intraoperative blood loss, operative time, and intraoperative radiation time and dose. However, some studies found increased operative and intraoperative radiation times and doses, as well as increased neurological complications. Patient-specific management and 3D printing demonstrated accurate hardware placement and a high-quality reduction without increasing the complication rate. These methods also decreased intraoperative blood loss and operative time. These new technologies also have some barriers to their use, the first of which is their additional cost. Using personalization tools takes time during the preoperative period. Second, they are mainly efficient for unilateral fractures, using the healthy side as a reference to reduce and fix the injured side. However, for bilaterally fractured pelvises where there is no healthy opposite side to use as a reference, another technique was developed to address the challenge. An average pelvic shape was created based on CT data from 40 individuals (111), suggesting that these average models support accurate and reliable virtual reduction reconstruction.

Author contributions:
MLB: Conceptualization, Writing - Original Draft, Writing - Review & Editing, Methodology, Data Curation
MP: Writing - Review & Editing, Supervision
SC: Writing - Review & Editing, Data Curation
XF: Conceptualization, Writing - Review & Editing, Supervision

Table 1. Results about RAFR on pelvic fractures

| Study | Patients | Fracture | Classification | Intraoperative blood loss | Operation time | Excellent and good reduction rate (Matta) | Postoperative complications | Global reduction between plan and postop result |
|---|---|---|---|---|---|---|---|---|
| Wu (22) | 20 | unstable pelvic fractures | 5 Tile B<br>15 Tile C | 133.25 ml | 204.75 min | 85% | 0 | |
| Zhao (9) | 15 | geriatric pelvic fractures | Rommens type 3 | 50 mL | 165 ±44 min | 100% | | 4.4 ±1.8 mm (1.8–8.5) |
| Zhao (17) | 22 | unstable pelvic fractures | 4 Tile B<br>18 Tile C | | | 95,50% | | 3.41mm ±1.83 |

Table 2. Results of RA in pelvic fractures

| Study | Patients | Fixation | Groups | Fracture | Intraoperative blood loss | Operation time | Fluoroscopy | Excellent and good reduction rate (Matta) | Postoperative complications | Screw placement | Majeed (last FU) |
|---|---|---|---|---|---|---|---|---|---|---|---|
| Wang (8) | 55 | all screws | RA 26 vs conventional 29 | pelvic fracture | 32 ± 18 vs 48 ± 51 (NS) | 187 ± 55 vs 231 ± 87 (S) | 13 ± 2 times vs 22 ± 7 (p<0.001) | | | | 94.8 ± 2.4 v 94.1 ± 2.4 (NS) |
| Zhao (24) | 64 | SI screws | 22 biplanar vs 21 RA vs 21 conventional | pelvic fracture | biplanar 12.76 ± 3.77 mL vs RA 11.92 ± 4.67 mL vs conventional 29.7 ± 8.01 mL (S) | 26.32 ± 6.32 min biplanar vs 79.24 ± 11.31 min conventional vs RA 15.81 ± 3.9 min (S) | biplanar 740.53 ± 185.91 cGy/cm2 vs RA 678.44 ± 127.16 cGy/cm2 (NS) vs conventional 2034.58 ± 494.54 cGy/cm2 (S) | | | biplanar 94.1% vs RA 97.2% vs conventional 75.7% (S) | |
| Zhu (6) | 119 | all screws | 24 nonop vs 45 ORIF vs 10 conventional vs 40 RA | non-displaced pelvic fractures | RA 27.5±16.7 mL vs conventional 32.9±21.4 mL vs ORIF 153.3±87.0 mL (both p<0.01) | | RA 62.6±30.1 times vs conventional 224.6±82.5 times (p<0.05) vs ORIF 2.2±1.4 times (p<0.01) | | 0 | FH grade 1, others =0 | nonop 90.7±4.1 vs conventional 91.8±3.8 vs RA 94.15±3.7 vs ORIF 88.6±4.1 |
| Liu (23) | 86 | all screws | Non-comparative | unstable pelvic fractures | | 175 min (35–280 min) | 29.1 times (9–63 times) | | | 13 unsatisfactory guiding needle placements | 92.4 |
| Liu (25) | 45 | robot-assisted SI screws + robot-assisted INFIX | 24 robot-assisted vs 21 conventional | | 35.0 ± 7.2 vs 46.2 ± 9.3 (p< 0.001) | 65.4 ± 10.9 vs 86.7 ± 14.7 (p< 0.001) | 29.2 ± 7.6 vs 52.3 ± 12.4 (p< 0.001) | | | | 86.4 ± 7.2 vs 84.3 ± 10.3 (NS) |
| Gilani (26) | 21 | IS or TS screws | 10 robot-assisted vs 11 conventional | pelvic fracture | | | | | | 4 violations (8%) vs 17 (31%) P = 0.02 | |
| Xing (27) | 46 | all screws | 24 robot-assisted vs 22 conventional | geriatric pelvic fracture | 17.04 +/- 3.14 vs 36.22 +/- 3.25 (p<0.001) | 34.1 ± 2.67 min vs 64.5 ± 4.19 min (p < 0.001) | | 91.67% versus 72.73%, (p < 0.001) | 0 | 100% vs 85.7% (S) | 81.09 ± 2.78 vs 71.27 ± 4.44 (p<0.001) |
| Li (28) | 57 | SI screws | 30 robot-assisted vs 27 conventional | pelvic fracture | | 34.503 ± 4.7677 vs 74.707 ± 2.4598 (p=0.000) | 197.995 ± 53.155cGycm2 vs 247.355 ± 32.314cGycm2 (p=0.000) | Matta points 22/8/0/0 vs 16/8/3/0 (NS) | | modified Gertzbein-Robbins classification 92.59% vs 78.57% (χ2 = 3.967, p< 0.05) | 25/5/0/0 vs 22/5/0/0 (NS) |
| Han (29) | 63 | S2 IS screws | 38 robot vs 25 conventional | unstable pelvic fractures | | | 8.06 V 3.54 s vs 27.37 ± 8.82 s (p< 0.001) | | | excellent and good rate 100% vs 85.7% (p< 0.001) | |
| Long (30) | 91 | IS screws | 56 robot / 35 conventional | unstable pelvic fractures | 33.89 ± 16.4 mL (15–80 ) vs 43.04 ± 12.34 mL (30–80) (p= 0.000) | 33.25 ± 6.46min (19–51) vs 63.55 ± 6.62 min (50–83) (p= 0.000) | 5.88 ± 1.29 min (3.3–- 8.2) vs 11.05 ± 2.98 min (5.9–17.3) (p= 0.000) | 89.28% vs 88.57% (NS) | | | 91.07% vs 91.43% (NS) |

S: significant
NS: non significant
nonop: non operative
ORIF: open reduction and internal fixation

Table 3. Results of navigation in pelvic fractures

| Study | Type | Patients | Fixation | Groups | Fracture | Intraoperative blood loss | Operation time | Fluoroscopy | Excellent and good reduction rate (Matta) | Postoperative complications | Screw placement |
|---|---|---|---|---|---|---|---|---|---|---|---|
| Florio (47) | non-comparative | 10 | | | | | | 877.8 mGy/cm2 (162.3-2040.16) | | | 2 screws perforation less than 2mm |
| Teo (48) | comparative | 36 (30,6% dysmorphic sacrum) | SI screws | 17 nav vs 19 conventional | | | | | | | 3 cortical breaches (grade 1), breach/dysmorphism (NS) breach/navigation (NS) |
| Madeja (50) | comparative | 57 | SI screws | 32 nav vs 25 conventional | | | 41 vs 45 min (S) | 650 vs 1024 cGy/cm2 (S) | | | 2 nav vs 15 misplacement |
| Mu (49) | non-comparative | 8 | perc trans-symphyseal screws | | | 10mL | 57 mn (30–80) | 40 s (36–47) | | | |
| Zwingmann (51) | prospective comparative | 187 | | 39 nv vs 83 | SI dislocation | | | | | | 4 vs 8 (intraoperative), 13 vs 29 (postoperative) (NS) |
| Zwingmann (51) | prospective comparative | 334 | | 75 nav vs 156 | sacral fracture | | | | | | 6 vs 6 (intraoperative), 17 vs 41 (postoperative) (NS) |
| Peng (52) | non-comparative | 13 | SI screws | | LC2-3 VS | | 21.2 ± 10.6 (10-45) | 26.4 1.5 mGy (22.5-26.8) | | 0 | |
| Ciolli (53) | non-comparative | 24 | | | | | 41 mn (±12.5) | | | | 2 grade1 |
| Boudissa (54) | comparative | 127 | | 45 nav vs 129 | | | 28.2 min vs. 21.6 min, (p=0.003) | 6.6 Gy/cm2 vs. 4.9 Gy/cm2 (p=0.02) | | | misplacement 8.5%, 11/129 vs. 2.2%, 1/45, respectively (p=0.19) |
| Wong (55) | non-comparative | 17 | all screws | | FFP | 94.1 ± 26.6 mL | | | | 1 infection | |

S: significant
NS: non significant
nav: navigation
FFP: fragility fracture of the pelvis

Table 4. Results of navigation in acetabular fractures

| Study | Type | Patients | Fixation | Groups | Fracture | Intraoperative blood loss | Operation time | Fluoroscopy | Complication | Screw placement |
|---|---|---|---|---|---|---|---|---|---|---|
| He (59) | comparative | 22 | perc AC retrograde screw | 10 navigation vs 12 conventional | AC or transverse | 22.0 ± 4.8 mL vs 45.0 ± 13.9mL ($p < 0.001$) | | 26.9 ± 2.1 s vs 80.7 ± 3.0 s ($p < 0.001$) | | 0/10 vs 2/12 misplacement |
| Lin (61) | prospective non comparative | 3 | perc AC retrograde screw | | non-displaced fractures | <10 ml | 40 min (30– 45) | 38 s (35–45 s) | | |
| Hong (60) | non-comparative | 18 | 21 AC screws and 9 PC screws | | | | 24.6 min (16–47) | 28.4 s (11–58) | 1 transient femoral nerve palsy | average deviated distance of wire tip 2.5mm (1.1–3.6) average trajectory difference 2.45° (1.5°–4.6°) |

AC: anterior column
PC: posterior column
Perc: percutaneous

Table 5. Results of patient specific management of acetabular fractures

| Study | Type | Patients | Fixation | Groups | Fracture | Intraoperative blood loss | Operation time | Fluoroscopy | Reduction | Postoperative complications | Screw placement |
|---|---|---|---|---|---|---|---|---|---|---|---|
| Maini (73) | prospective randomized case-control study | 21 | reduction of the fracture was done on the 3D real model and plate contouring pre-operatively | 10 PS vs 11 | acetabulum | 100 ml less in PS | 12 min less in PS | | 0 PS vs 4 step of 2–3 mm (S) | | |
| Sakong (74) | comparative | 41 | precontouring on 3D model obtained by mirroring | 9 PS vs 11 | acetabulum | NS | instrumentation 40.6 vs 91 min (S), operative NS | | | | |
| Li (75) | comparative | 16 | reduction of fracture on the simulation software, pre-contouring plate on 3D model mirrored on the healthy side | 7 PS vs 9 | acetabulum + hip dislocation | NS | fixation -33 min (S), operative -43 min (S) | | NS | 2/7 vs 5/9 | |
| Ijpma (76) | non-comparative | 10 | 3D virtual model, planned PS printed plate and screw trajectory, 3D printed guides | | displaced associated-type acetabular | | | | median gap was reduced to 3 (2-5) mm (p=0.005), median step-off was reduced to 0 (0-2) mm (p=0.01) | | median difference in screw direction between the planning and actual surgery 7.1° (7°-8°) |
| Agarwal (71) | non-comparative | 30 | plate model designed, printed ad the plate is precontoured based on the model | | acetabulum, KL | 367.85 ml simple F vs associated F 450 ml (S) | 89.8 (sd = 22.2) min | | | | |

S: significant
NS: non significant
KL: Kocher Langenbeck
PS: patient specific

Table 6. Results of reduction and plate pre-contouring on a 3D model for acetabular fractures

| Study | Type | Patients | Method | Groups | Fracture | Printing time | Intraoperative blood loss | Operation time | Fluoroscopy | Reduction | Postoperative complications |
|---|---|---|---|---|---|---|---|---|---|---|---|
| Wang (95) | randomized comparative | 48 | ORIF 3D printing assisted for preoperative planning vs traditional | 24 vs 24 | acetabulum | | 1132.80 ± 70.84 mL vs 1401.62 ± 43.37 mL (S) | 152.03 ± 29.09 vs 187.53 ± 18.85 mn (S) | | 16 excellent, 4 good, 4 fair vs 8 excellent, 4 good, 12 fair (S) | |
| Weidert (96) | non-comparative | 12 | reduction on the 3D printed model (with a newly introduced surface filtering method) | | acetabulum | 8h40 vs 25h26 without the surface filter | 853mL | 3h16 | | | |
| Chen (97) | comparative | 52 | | 28 CAS + 3D printing precontoured plate vs 24 | acetabulum, ABC | | 696.07 ± 166.54 mL vs 833.75 ± 227.44 mL (p=0.018) | 157.5 ± 20.48 min vs 187.08 ± 35.81 min (p=0.001) | | excellent 25/28 v 18/24 (NS) | |
| Wan (98) | comparative | 96 | | 48 CAS reduction +3D vs 48 | acetabulum | | 1,147.2±235.4 vs 1,832.5±268.1 (p<0.001) | 210.8±54.5 vs 296.4±66.2 (p<0.001) | 6.8±1.6 vs 12.4±2.1 (p<0.001) | excellent / good 81.25 % vs 77.08 | 5/48 vs 18/48 (p=0.002) |
| Ivanov (99) | comparative | 22 | | 12 3D vs 10 | acetabulum | | 665 +/- 52.7 vs 837.5 +/- 68 (p<0.01) | 193.5 +/- 26 vs 252.5 +/- 30.2 (p<0.01) | 25.9 +/- 5.6 vs 66.5 +/- 6.7 (p<0.01) | 80% good vs 50% | |
| Ansari (100) | comparative | 27 | | 3D 12 vs 15 | acetabulum | 728 ± 32.3 min | 664 ± 186.4 ml vs 936 ± 198.4 ml (S) | 184.4 ± 39.2 min vs 246.4 ± 44.2 min (S) | 22 ± 5.6 shots vs 62 ± 16.5 (S) | NS | |
| Huang (101) | randomized comparative | 40 | plate prebending | 20 3D + preoperative reduction vs 20 | acetabulum | | 500 [400, 800] mL (p<0.001) | operation 130.8 ± 29.2 min, t = 7.5 (p<0.001) instrumentation 32.1 ± 9.5 min, t = 6.5 (p<0.001) | 4.2 ± 1.8 vs. 7.7 ± 2.6 s P < 0.001 | good reduction 80% vs. 30% (S) | 5% vs 25% (NS) |
| Li (75) | comparative | 16 | virtual simulation + 3D printing + precontoured plate | 7 vs 9 | acetabulum + hip dislocation | | (NS) | fxation -33 min (S) operative -43mn (S) | | (NS) | 2/7 vs 5/9 |
| Zeng (82) | non-comparative | 10 | 3D printing + CAS (virtual reduction, virtual positioning of hardware, 3D printing, pre-contouring on solid 3D model, verification with postop CT) | | | | | | | 7/10 less than 1 mm, 3/10 1-2 mm | |
| Chana Rodriguez (78) | non-comparative | | pre-contouring plates on 3D printed models for acetabular fractures | | acetabulum | | | intrumentation 16.9 min (10-24) | | | |

S: significant
NS: non significant
ABC: associated both columns fracture

Table 7. Results of 3D printing of a template for positioning hardware in pelvic and acetabular fractures

| Study | Type | Patients | Method | Groups | Fracture | Printing time | Intraoperative blood loss | Operation time | Fluoroscopy | Reduction | Screw placement |
|---|---|---|---|---|---|---|---|---|---|---|---|
| Liu (91) | non-comparative | 17 | template for insertion of SI screw | | unstable pelvic fractures | | each screw 32.0 ± 11.1 ml (20–50 ml) | each screw 45.9 ± 8.6 min (30–60 min) | each screw 50.3 ± 19.7s (24–96 s) | 17 excellent | 18 Grade 1, 2 Grade 2 |
| wu (92) | comparative | 37 | virtual reduction, virtual placement of screw, design of a template to guide screw placement | 42 screws template group vs 31 | pelvic fractures | | | each screw 25.01 ± 2.90 min vs 46.24 ± 9.59 min (S) | each screw 12.05 ± 4.00 vs 56.10 ± 6.75 | | perforation 2/42 vs 5/38 (S) |
| wu (90) | non-comparative | 50 | template for anterograde transpubic screw | template = 9 | pelvic fractures | | 141.11 ± 50.21 ml | 108.75 ± 25.71 min | 17.89 ± 4.01 times | | |
| Chen (93) | non-comparative | 64 plates 339 screws | printed template | | acetabulum | | | | | | designed and real trajectories (NS) |
| Ijpma (76) | non-comparative | 10 | 3D virtual model, planned patient specific printed plate and screw trajectory, 3D printed guides | | displaced associated-type acetabular | | | | median gap was reduced to 3 (2-5) mm (p=0.005) median step-off was reduced to 0 (0-2) mm (p=0.01) | | median difference in screw direction between the planning and actual surgery 7.1° (7°-8°) |

S: significant
NS: non significant

**Figure legends:**

Figure 1: flow chart